\documentclass[sigconf]{acmart}
\usepackage{graphicx}
\usepackage{subcaption}

\usepackage{booktabs} 
\usepackage{amsmath}

\setcopyright{rightsretained}

\renewcommand\footnotetextcopyrightpermission[1]{}

\acmDOI{}

\acmISBN{}

\acmConference[PCSC2025]{Philippine Computing Science Congress}{May 2025}{Benguet, Philippines}
\acmYear{2025}
\copyrightyear{2025}

\acmArticle{}
\acmPrice{}

\editor{}
\editor{}
\editor{}

\begin{document}
\title{Representing Classical Compositions through \\ Implication-Realization Temporal-Gestalt Graphs}
\subtitle{An Exploratory Study into Graph-Based Musical Representation}





\author{Al Vincent Bomediano}
\affiliation{%
  \institution{Ateneo de Manila University}
  \streetaddress{Katipunan Ave}
  \city{Quezon City}
  \state{Metro Manila}
  \postcode{1108}
}
\email{al.bomediano@student.ateneo.edu}

\author{Raul Jarod Conanan}
\affiliation{%
  \institution{Ateneo de Manila University}
  \streetaddress{Katipunan Ave}
  \city{Quezon City}
  \state{Metro Manila}
  \postcode{1108}
}
\email{raul.conanan@student.ateneo.edu}

\author{Lance Dominic Santuyo}
\affiliation{%
  \institution{Ateneo de Manila University}
  \streetaddress{Katipunan Ave}
  \city{Quezon City}
  \state{Metro Manila}
  \postcode{1108}
}
\email{lance.santuyo@student.ateneo.edu}

\author{Andrei Coronel, PhD}
\affiliation{%
  \institution{Ateneo de Manila University}
  \streetaddress{Katipunan Ave}
  \city{Quezon City}
  \state{Metro Manila}
  \postcode{1108}
}
\email{acoronel@ateneo.edu}

\renewcommand{\shortauthors}{}

\begin{abstract}
Understanding the structural and cognitive underpinnings of musical compositions remains a key challenge in music theory and computational musicology. While traditional methods focus on harmony and rhythm, cognitive models such as the Implication-Realization (I-R) model and Temporal Gestalt theory offer insight into how listeners perceive and anticipate musical structure. This study presents a graph-based computational approach that operationalizes these models by segmenting melodies into perceptual units and annotating them with I-R patterns. These segments are compared using Dynamic Time Warping (DTW) and organized into k-nearest neighbors (k-NN) graphs to model intra- and inter-segment relationships.

Each segment is represented as a node in the graph, and nodes are further labeled with melodic expectancy values derived from Schellenberg’s two-factor I-R model—quantifying pitch proximity and pitch reversal at the segment level. This labeling enables the graphs to encode both structural and cognitive information, reflecting how listeners experience musical tension and resolution.

To evaluate the expressiveness of these graphs, we apply the Weisfeiler-Lehman (WL) graph kernel to measure similarity between and within compositions. Results reveal statistically significant distinctions between intra- and inter-graph structures. Segment-level analysis via multidimensional scaling (MDS) confirms that structural similarity at the graph level reflects perceptual similarity at the segment level. Graph2vec embeddings and clustering further demonstrate that these representations capture stylistic and structural features—such as texture, phrasing, and melodic contour—that extend beyond composer identity.

These findings highlight the potential of graph-based methods as a structured, cognitively informed framework for computational music analysis, enabling a more nuanced understanding of musical structure and style through the lens of listener perception.

\end{abstract}

\ccsdesc[500]{Applied computing~Sound and music computing}
\ccsdesc[500]{Mathematics of computing~Graph algorithms}
\ccsdesc[300]{Theory of computation~Kernel methods}
\ccsdesc[300]{Information systems~Similarity measures}
\ccsdesc[300]{Information systems~Clustering and classification}

\keywords{Music analysis, Graph-based representation, k-NN graph, Implication-Realization model, Dynamic Time Warping, Graph kernels, Graph embeddings, k-means clustering}

\maketitle

\section{Introduction}

Music theory seeks to understand the principles governing musical composition and perception. While traditional approaches focus on the structural elements of music—such as harmony, rhythm, and counterpoint—music cognition shifts attention to the listener’s experience. 

The Implication-Realization (I-R) Model by Narmour provides a cognitive framework on melodic expectation, suggesting that listeners form innate expectations about melodic direction as the melody unfolds \cite{Foster1990}. Complementing this, Temporal Gestalt Perception theory describes how human auditory perception segments continuous musical stimuli into discrete, meaningful units through hierarchical organization in which basic elements combine to form increasingly larger structures when grouped together \cite{Tenney1980}.

These models have seen limited computational adoption. Existing graph-based approaches to music analysis primarily model compositions through established musicological frameworks—such as Schenkerian analysis and the Generative Theory of Tonal Music (GTTM)—using hierarchical trees to represent musical structure \cite{Simonetta2018, Orio2009}, while others employ statistical methods to identify structural patterns \cite{Zou2022}.

This study introduces a graph-based computational model that represents melodic expectation and musical perception as described by the I-R Model and Temporal Gestalt Perception theory. The proposed method applies segmentation techniques from prior work \cite{Simonetta2018, Orio2009} and constructs similarity-based graphs using Dynamic Time Warping (DTW) \cite{JMLR:v21:20-091} distance to compare musical segments. 

Graph kernels \cite{JMLR:v21:18-370} are used to evaluate intra- and inter-graph similarity, with the expectation that self-similarity will be higher than cross-composition similarity—indicating meaningful structural representation of distinctive musical features. To further validate the expressive power of this representation, graphs are embedded using graph2vec \cite{narayanan2017graph2vec} and the resulting vectors are clustered to demonstrate that compositions can be differentiated based on their graph-encoded features.


\section{Related Work}

\subsection{The Implication-Realization Model}

The Implication-Realization (IR) Model, developed by Narmour, provides a bottom-up framework for melodic expectation based by emphasizing the fundamental relationships between successive notes \cite{Royal1995, Foster1990}. Each pattern reflects a listener’s expectation for continuity or change, categorized into archetypes like Process (P), Duplication (D), and Reversal (R). These categories depend on interval size and direction, shaping how listeners perceive closure, tension, or surprise.

Narmour formalized this with two basic hypotheses:
\begin{align*}
    A + A \rightarrow A \\
    A + B \rightarrow C
\end{align*}
These suggest that repetition suggests further repetition while change suggests further change. Together, they describe how listeners anticipate the continuation of a melody \cite{Foster1990}.


\subsubsection{\textbf{Rule-Based I-R Symbol Assignment}}
IR symbol labeling is implemented using Wen et al.’s interval-based encoding \cite{8407293}, extended by Noto et al.’s rule-based method \cite{noto2021rule_based}, which assigns IR symbols based on interval size and direction with a fixed threshold for small and large intervals.

\subsubsection{\textbf{Two-Factor I-R Model}}
We adopt Schellenberg's simplified two-factor model of melodic expectancy, demonstrated to accurately capture listeners’ expectancy judgments while reducing conceptual redundancy \cite{Schellenberg1997}. The squared semipartial correlations (sr² = 0.364 for Proximity, 0.144 for Reversal) from an experiment on Webern lieder will be used to match our classical corpus.

\subsection{Temporal Gestalt Units and Musical Segmentation}
Tenney and Polansky’s Temporal Gestalt Unit (TG) theory \cite{Tenney1980} describes how listeners segment continuous musical input into discrete, meaningful units. Unlike traditional models that treat music as a linear sequence of notes, TG theory posits that human auditory perception actively segments the continuous flow of musical stimuli into discrete, meaningful units.

The theory outlines a hierarchical organization of musical elements. At the base are elements (pitches, durations, timbres), which combine into clangs—brief perceptual units akin to small musical gestures. At a higher level, sequences organize clangs into ordered patterns, establishing structural coherence and directional flow within a TG. The highest level of the hierarchy consists of segments, which integrate clangs and sequences into extended musical phrases or motifs. These segments contribute to overarching musical structures, incorporating variations while maintaining perceptual continuity.

\subsection{Dynamic Time Warping}
Dynamic Time Warping (DTW) is a time series alignment method that enables non-linear matching between sequences of unequal lengths by minimizing cumulative distance through a flexible warping path \cite{sakoe1978dynamic}. Its multivariate extension aggregates distances across multiple features \cite{JMLR:v21:20-091}, making it well-suited for comparing musical segments represented by pitch, duration, and I-R patterns.

DTW is particularly appropriate in this context due to the variability introduced by Gestalt-based segmentation. It captures structural similarity even when segments differ in rhythmic density, timing, or ornamentation—aligning stretched or compressed phrases that share similar melodic contour. This flexibility mirrors how listeners perceive equivalence across expressive variations, making DTW a perceptually grounded and structurally robust distance metric for music analysis.

\subsection{Graph Kernels and Similarity Computation}
Graph kernels assess similarity by applying the kernel trick to compare graphs in an implicit high-dimensional space, enabling efficient structural comparison without explicit feature extraction \cite{NIPS2000_4e87337f}. While widely used in machine learning (e.g., SVMs, PCA) \cite{scholkopf1999advances}, their use in graph analysis allows for the comparison of complex structures that traditional methods struggle to represent. A range of graph kernels exist, each capturing different aspects of graph topology, making them essential for graph-based machine learning tasks \cite{ediss7169}. 

Among these, the Weisfeiler-Lehman kernel is particularly appropriate for this study, as it captures both local and hierarchical graph structures through iterative neighborhood aggregation. This is especially effective for the segment-based graphs, where musical structure is encoded in localized patterns of connectivity \cite{JMLR:v12:shervashidze11a}.


\subsection{Graph Embeddings and Clustering}
Many graph analysis tasks require fixed-length vector representations to enable the use of conventional machine learning models. While graph kernels can assess structural similarity, they often lack explicit embeddings. Graph2Vec addresses this by learning unsupervised, fixed-size embeddings for arbitrary graphs. Inspired by Doc2Vec, it treats graphs as documents and rooted subgraphs as words, using Weisfeiler-Lehman relabeling and a skipgram-based model to preserve structural features. This results in fixed-dimensional embeddings where similar graphs are placed closer together in the embedding space, allowing for graph analytics tasks such as clustering and classification in a data-driven and scalable manner. \cite{narayanan2017graph2vec}.

\section{Methodology}


\begin{figure}[]
    \centering
    \includegraphics[width=0.5\linewidth]{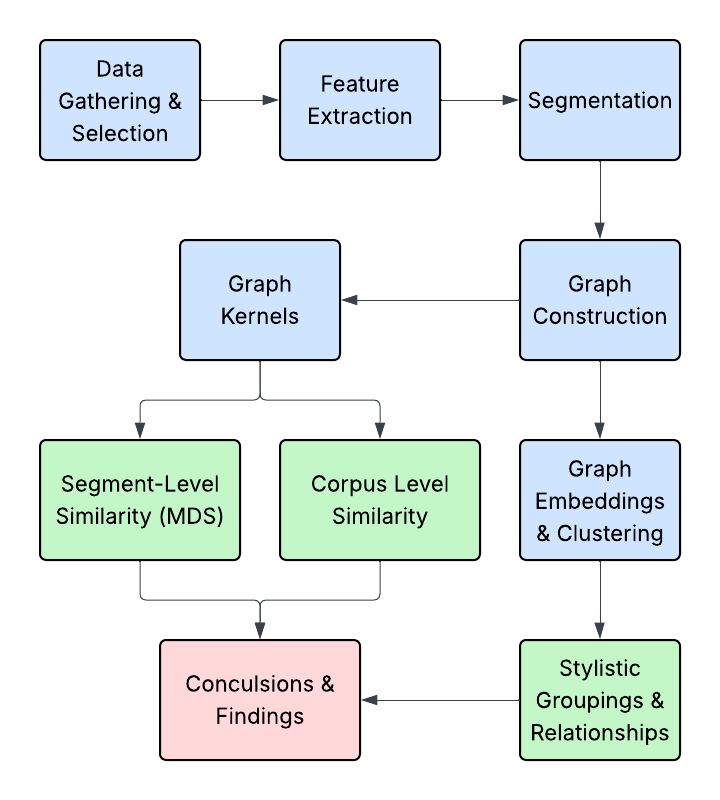}
    \caption{Methodology Flowchart}
    \label{fig:method-flow}
\end{figure}

\subsection{Data Gathering \& Selection}
MusicXML transcriptions of works by composers such as Bach, Chopin, and Ysaÿe were collected, focusing on pieces with clear melodic lines that remain intact without accompaniment. Using the music21 library, each file was converted into a structured note matrix for analysis.

The dataset includes Bach’s Cello Suites, Ysaÿe’s Violin Sonatas, and Chopin’s Études and selected Waltzes—all chosen for their emphasis on melodic development and variation, providing a strong foundation for graph-based similarity analysis and identity representation.


\subsection{Note Matrix Formation}
The note matrix is based on the implementation found in MIDIToolbox \cite{miditoolbox2016} but is adapted to align with the specific analytical needs of this study. Each row represents a note, while columns correspond to the following selected musical features:
\begin{itemize}
  \item Pitch \& Octave: Captures the tonal characteristics of each note.
  \item Duration \& Onset: Represents the rhythmic placement of each note within the composition.
  \item Beat Strength: Measures the metric weight of of each note within its measure.
\end{itemize}

\subsubsection{\textbf{Implication-Realization Pattern Assignment}} 

\begin{figure}[h]
    \centering
    \includegraphics[width=\linewidth]{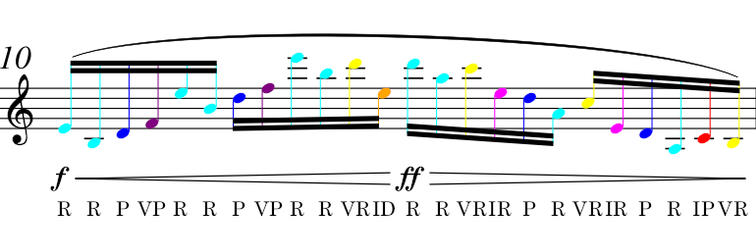}
    \caption{An excerpt from Chopin’s \textit{Étude Op. 25 No. 11} ("Winter Wind"), annotated using the Implication-Realization (IR) model. Notes are color-coded and labeled to illustrate IR patterns.}
    \label{fig:ww_snippet}
\end{figure}

To encode information about melodic expectancy, each note is annotated with an Implication-Realization (I-R) pattern based on Narmour's I-R Model. The assignment of I-R symbols to each triplet of notes is adapted from the rule-based method described by Noto et al. \cite{noto2021rule_based}, ensuring an accurate and complete mapping of symbols based on Narmour's model. Figure~\ref{fig:ww_snippet} provides a visual representation of this annotation.

The I-R symbol is added as a column in the note matrix, each capturing Narmour's core expectancy principles of registral direction, intervallic difference, registral return, proximity, and closure. 

\begin{table}[h]
    \centering
    \small
    \renewcommand{\arraystretch}{1.1} 
    \setlength{\tabcolsep}{2pt} 
    \begin{tabular}{ccccccccc}
        \hline
        Onset & Onset & Dur. & MIDI & PC & Octave & Beat & IR  \\
        (global) & (measure) & (beats) & Pitch & Class &  & Strength & Symbol  \\
        \hline
        0.00 & 0.00 & 0.50 & 66 & 6 & 4 & 0.250 & X \\
        0.50 & 0.50 & 1.50 & 66 & 6 & 4 & 1.000 & DP \\
        2.00 & 2.00 & 1.00 & 66 & 6 & 4 & 0.500 & P \\
        3.00 & 3.00 & 0.25 & 65 & 5 & 4 & 0.250 & P \\
        3.25 & 3.25 & 0.25 & 63 & 3 & 4 & 0.125 & VR \\
        3.50 & 3.50 & 1.50 & 70 & 10 & 4 & 1.000 & R \\
        \hline
    \end{tabular}
    \caption{A tabular representation of a note matrix, where each row corresponds to a note and each column represents a musical feature.}
    \label{tab:note_matrix}
\end{table}

This structured representation of the note matrix ensures the preservation of essential symbolic and perceptual information while allowing for computational processing and analysis.


\subsection{Temporal Gestalt Segmentation}
Segmentation is guided by Gestalt principles, grouping notes into meaningful units based on proximity in pitch and duration—reflecting cognitive theories of musical perception \cite{Tenney1980}. The implementation follows the method used in MIDI Toolbox \cite{miditoolbox2016}.
\begin{figure}[h]
    \centering
    \includegraphics[width=\linewidth]{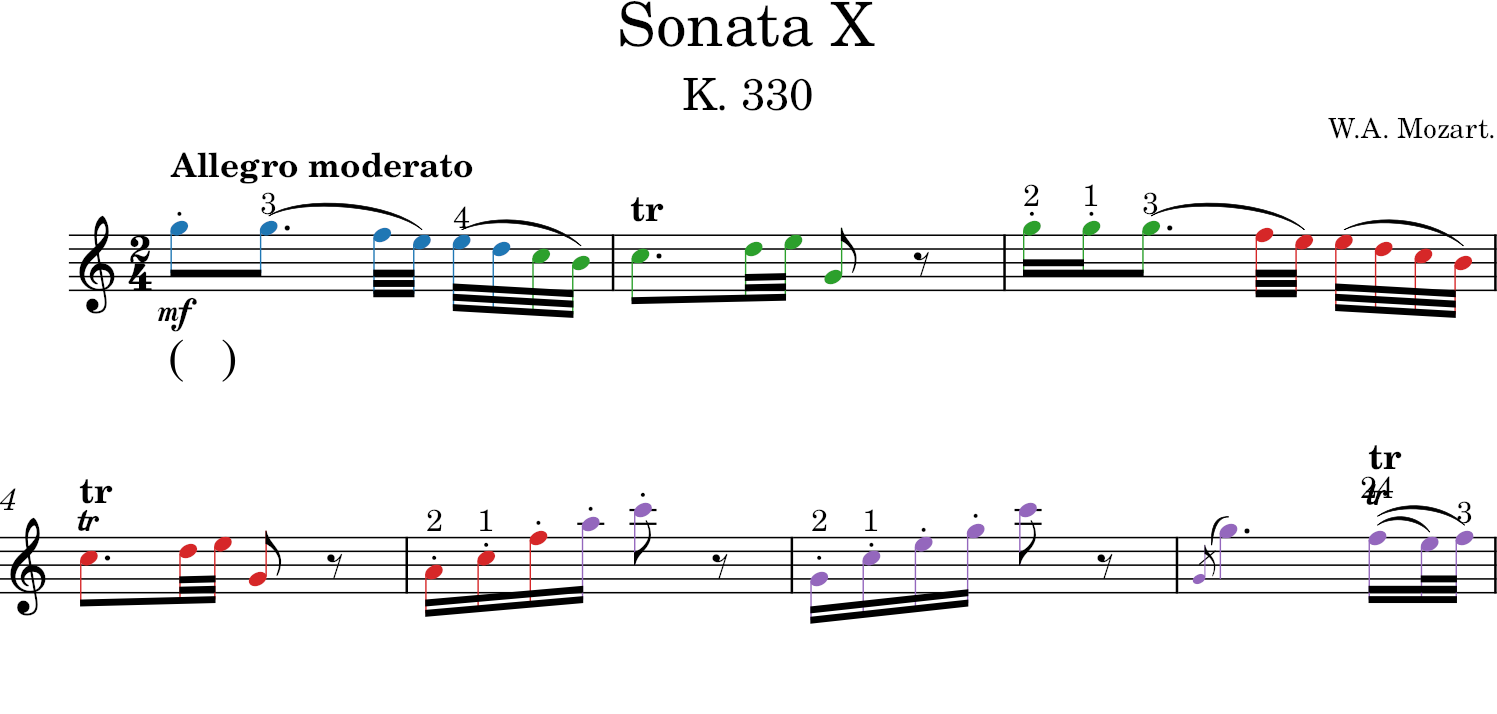}
    \caption{An excerpt from the first movement of Mozart’s \textit{Sonata No.\ 10 in C major (K.\ 330)}, illustrating the segmentation approach. Each color-coded group of notes indicates a distinct segment identified through Gestalt-based rules.}
    \label{fig:mz_sg1}
\end{figure}

Each resulting segment serves as a node in the graph representation, encapsulating perceptually coherent musical ideas.

\subsection{Graph Construction}
The graph representation is defined as \( G = (V, E) \), where nodes represent musical segments, and edges are weighted based on the Dynamic Time Warping (DTW) distances between segments.
\subsubsection{\textbf{Graph Definition}}
The adjacency matrix of the graph, also referred to as the distance matrix, is defined as:
\begin{equation}
    A_{i,j} = DTW(S_i,  S_j)
\end{equation}
where \( S_i \) and \( S_j \) are two different musical segments and \( DTW(S_i, S_j) \) computes the optimal alignment cost between their note matrices. 


\subsubsection{\textbf{k-Nearest Neighbors Graph Construction}}
To transform the distance matrix into a graph structure, a k-nearest neighbors (k-NN) approach is applied. Each node is connected to its \(k\) most similar neighbors, with similarity defined as the inverse of the DTW distance.

\begin{figure}[h]
    \centering
    \includegraphics[width=0.5\linewidth]{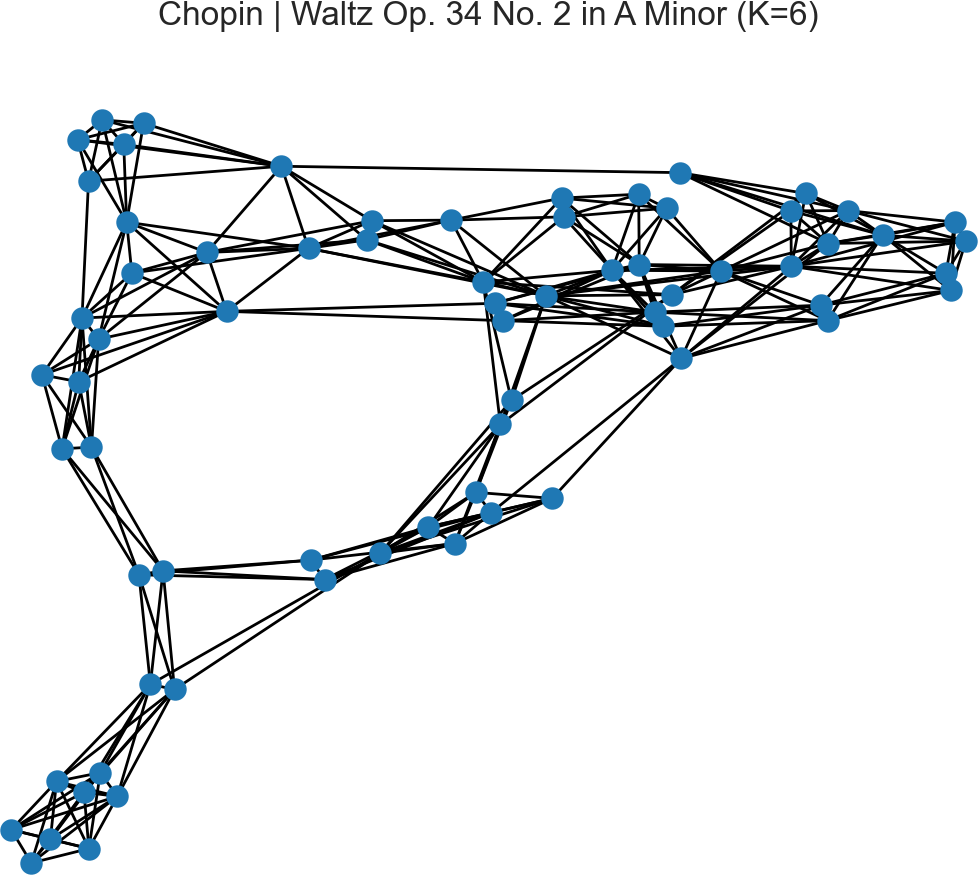}
    \caption{A graph representation of Chopin's \textit{Waltz Op. 34 No. 2 in A Minor}.}
    \label{fig:chopin_waltz}
\end{figure}

\subsubsection{\textbf{Node Labels}}

Each node in the graph is assigned a qualitative label, as required by categorical graph kernels such as the Weisfeiler-Lehman subtree kernel. Rather than using raw continuous expectancy scores, we convert them into discrete bins to form categorical node labels that reflect local perceptual structure.

The labeling scheme combines two cognitively salient features derived from the I-R model:

\paragraph{\textbf{Binned Expectancy Score}}
This expectancy value is derived from note-level scores, calculated according to Schellenberg’s two-factor simplification of the I-R model \cite{Schellenberg1997}.

For each note \(i\), an expectancy score \(E_i\) is computed based on its relationship with the preceding and following notes. The score combines two factors: Pitch Proximity \((PP)\) and Pitch Reversal \((PR)\), which measure the degree to which two consecutive intervals influence expectancy. Each factor is normalized and weighted according to Schellenberg’s empirical coefficients:
\[
E_i = \beta_{PP} \times PP_{norm} + \beta_{PR} \times PR_{norm}
\]

To obtain the standardized linear weights \((\beta)\), we use the semipartial correlation values \((sr^2)\) reported in Schellenberg’s Experiment 2 (Webern lieder corpus), taking the signed square root of each:
\[
\beta_{PP} = \sqrt{0.364} \approx 0.604, \qquad \beta_{PR} = \sqrt{0.144} \approx 0.379
\]
To obtain a single expectancy score for a graph segment \(S_j\), we take the arithmetic mean of all note-level expectancy values within that segment:
\[
E(S_j) = \frac{1}{|S_j|} \sum_{i \in S_j} E_i
\]
These segment-level expectancy values are then binned into five discrete categories — \texttt{VeryLow}, \texttt{Low}, \texttt{Medium}, \texttt{High}, and \texttt{VeryHigh} — based on their percentile rank within the dataset. This binning transforms continuous expectancy into a categorical form suitable for label-sensitive graph kernels.

\paragraph{\textbf{Dominant I-R Symbol}}
The second component of the node label is the most frequent I-R symbol (e.g., \texttt{P} for Process, \texttt{D} for Duplication) observed within a segment. These symbols act as discrete perceptual categories, each encoding a distinct form of melodic expectation \cite{noto2021rule_based}.

Each node label thus takes the form \texttt{(ExpectancyBin|IRSymbol)}. For example, a node labeled \texttt{High|P} represents a segment with high melodic expectancy that follows a Process pattern — a common archetype associated with smooth, directional motion. This combined labeling scheme integrates the two most informative dimensions of the I-R model, potentially providing a richer and more musically meaningful signature for each segment.

\subsection{Optimizing Graph Identity}
To verify that the constructed graph representation effectively captures the unique melodic and structural characteristics of each composition, and to determine the optimal \(k\) for the constructed k-NN graphs that best preserve musical identity, graph similarity is evaluated. This is achieved by comparing intra-graph similarity (within a single graph) to inter-graph similarity (between graphs representing different compositions). 

\subsubsection{\textbf{Graph Kernels}}
To compute similarity scores, we employ the Weisfeiler-Lehman kernel which captures hierarchical and localized structural features by iteratively refining node labels based on neighborhood information.

\subsubsection{\textbf{Intra-graph Similarity}}
To assess whether a graph’s internal structure is cohesive and retains its unique musical features, each graph is partitioned into two subgraphs, and their similarity is evaluated.

The Kernighan-Lin algorithm \cite{kernighan1970efficient} is used for partitioning, as it iteratively swaps pairs of nodes to minimize the total cut cost, ensuring that:
\begin{itemize}
    \item The partitioning process preserves the structural relationships between musical segments.
    \item Each partition maintains balanced subgraph sizes, preventing significant asymmetry.
    \item The algorithm respects DTW-based segment similarity, ensuring that musically similar segments are not arbitrarily separated.
\end{itemize}
\begin{figure}[h]
    \centering
    \includegraphics[width=\linewidth]{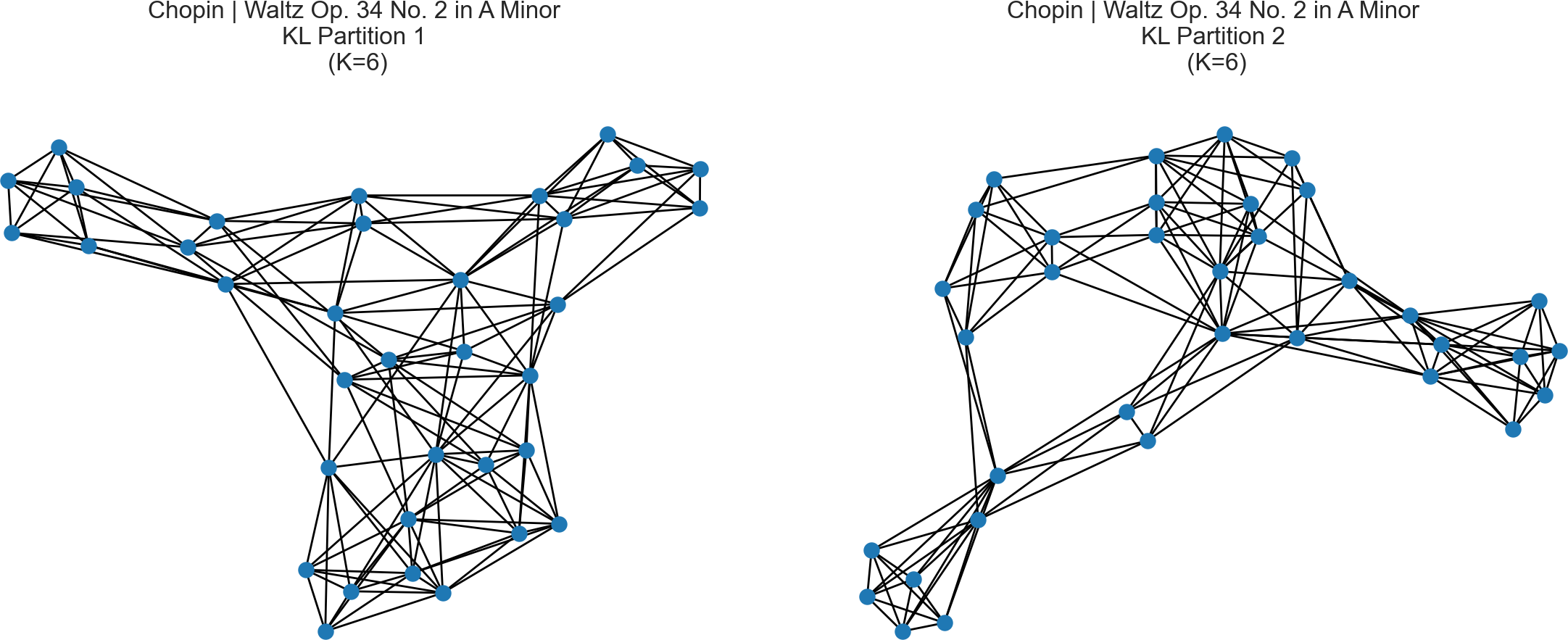}
    \caption{Partitions of Chopin's \textit{Waltz Op. 34 No. in A Minor} using the Kernighan-Lin algorithm.}
    \label{fig:waltz_kl}
\end{figure}

By systematically evaluating the consistency of each graph when divided, this approach provides a quantitative measure of its internal coherence.

\subsubsection{\textbf{Inter-graph Similarity}}
To evaluate the distinctiveness of each graph representation, similarity scores between I-R graphs of different compositions are compared. A lower similarity score across graphs, relative to intra-graph similarity, indicates that each graph preserves a distinct musical identity.

\subsubsection{\textbf{Comparing Intra-graph and Inter-graph Similarity Measures}}
For each \( k \)-level, intra-graph and inter-graph similarity scores are computed and statistically tested to determine whether the difference between the two groups is significant. This analysis is based on the premise that a graph representing a composition should exhibit greater similarity to itself than to graphs representing other compositions.  


\subsubsection{\textbf{Segment Level Similarity via MDS}}
To analyze structural similarity between compositions at the segment level, Multidimensional Scaling (MDS) is applied to select pairs of graphs. For each selected graph pair \(G_1, G_2\), we concatenate their segments into a combined list:
\[
M = \{s_1, ..., s_n \mid s \in G_1\} \cup \{t_1, ..., t_m \mid t \in G_2\}
\]
A joint distance matrix is then computed over all segments in \(M\) using Dynamic Time Warping (DTW) applied to the same multivariate features used during graph construction. This results in an \(|M| \times |M|\) matrix capturing pairwise segment distances across both pieces. We then apply MDS to project the high-dimensional distance matrix into a 2D space, allowing for visual inspection of how segments from the two pieces group. 


\subsection{Graph Embeddings and Clustering}
To further validate that the generated graphs effectively capture the unique melodic and structural characteristics of each composition, Graph2Vec \cite{narayanan2017graph2vec} is used to transform the graphs into fixed-size vector embeddings. These embeddings are then clustered using KMeans \cite{macqueen1967some}, with \( k \) set to the number of artists in the collected corpus. This approach assumes that while each graph represents a distinct composition, it should exhibit greater similarity to other graphs corresponding to compositions that share implicit and explicit musical features. Setting \( k \) to the number of artists is based on the expectation that compositions by the same artist—or those with similar stylistic and structural characteristics—will cluster together, demonstrating that the generated graphs effectively capture the unique musical identity of each composer.

After graph construction, each composition’s graph representation is processed through Graph2Vec to produce a fixed-length vector embedding. These embeddings are then clustered using KMeans, and the resulting labels are assigned to the corresponding compositions. To facilitate visualization, Principal Component Analysis (PCA) \cite{abdi2010principal} is applied for dimensionality reduction, allowing the clustering results to be represented in two dimensions.

\section{Results}
\subsection{Neighborhood‑size ($k$) optimization}

\begin{figure}[h]
    \centering
    \includegraphics[width=0.8\linewidth]{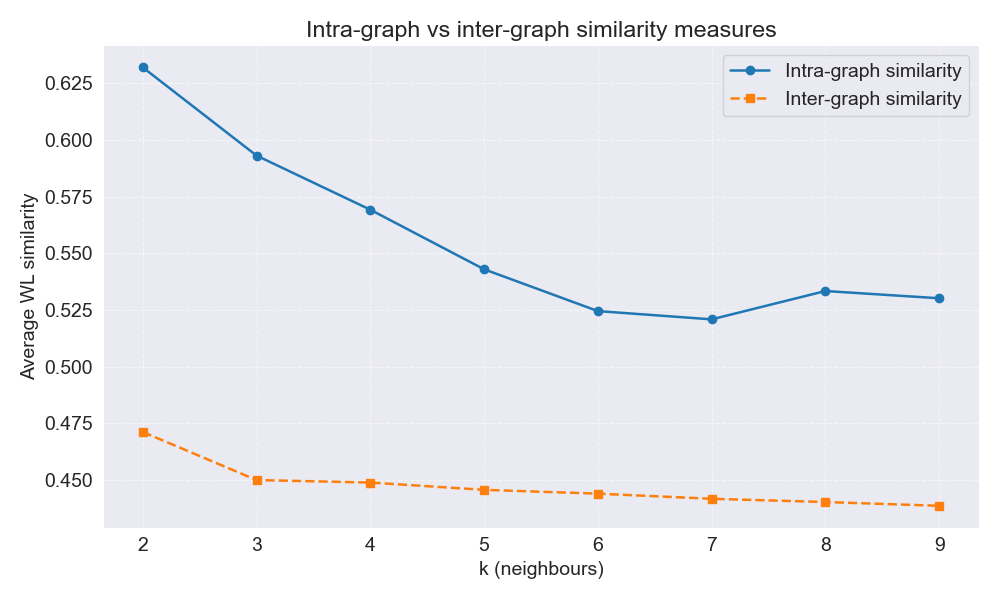}
    \caption{Mean intra‑ vs inter‑graph WL similarity across $k$}

    \label{fig:wl-kernel-optimization}
\end{figure}

\begin{figure}
    \centering
    \begin{subfigure}[b]{0.45\linewidth}
        \centering
        \includegraphics[width=\linewidth]{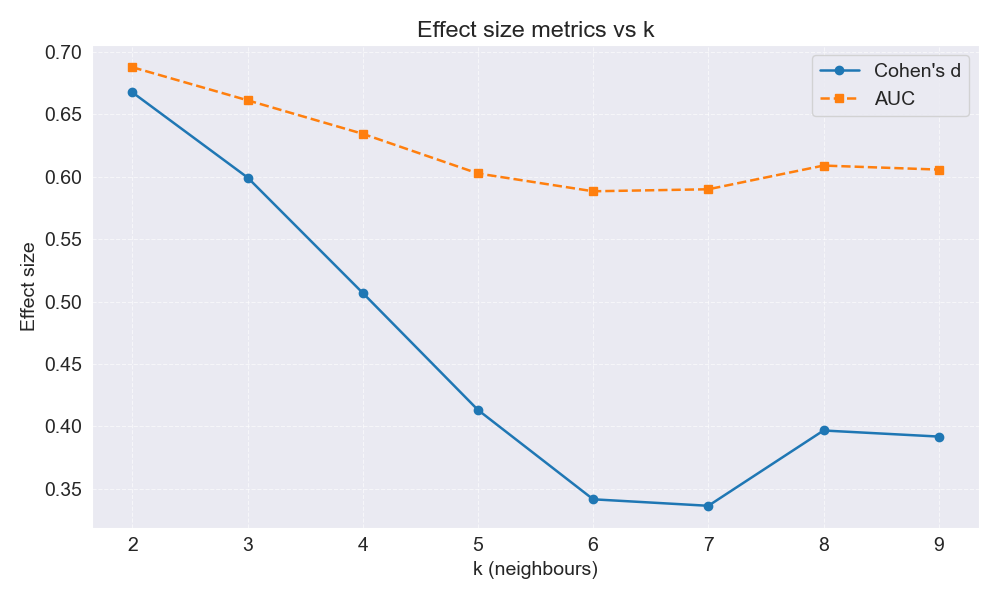}
        \caption{Effect‑size curves (Cohen’s $d$ \& AUC) vs $k$}
        \label{fig:kernel-stats_cohens}
    \end{subfigure}
    \hfill
    \begin{subfigure}[b]{0.45\linewidth}
        \centering
        \includegraphics[width=\linewidth]{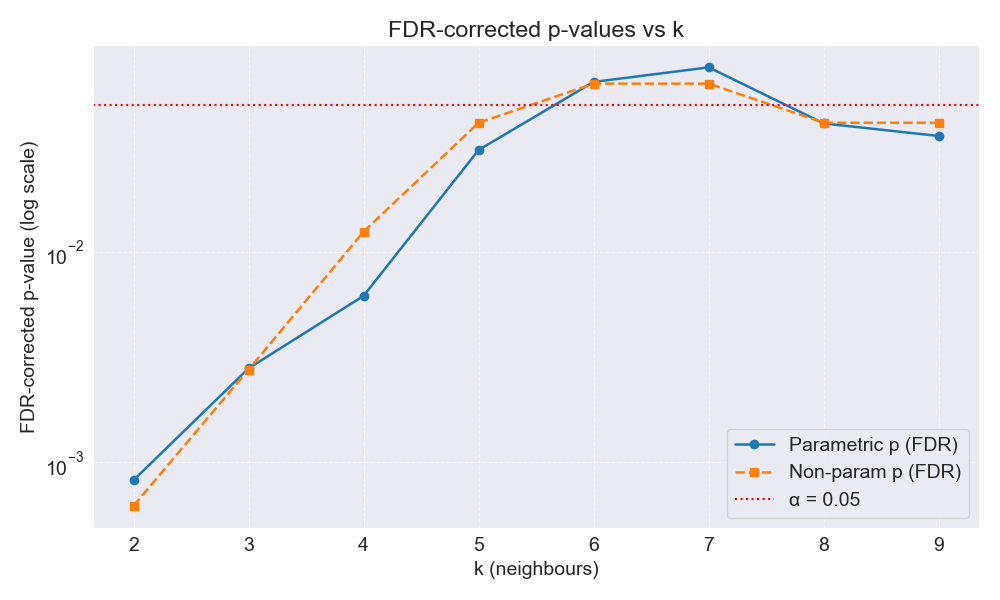}
        \caption{FDR‑corrected p-values on log‑scale for the normalised WL kernel}
        \label{fig:kernel-stats_p-values}
    \end{subfigure}
    \caption{Comparison of WL similarity statistics across $k$}
    \label{fig:kernel-statistics}
\end{figure}

Fig~\ref{fig:wl-kernel-optimization}  shows the expected monotonic decay in intra‑graph similarity as extra edges blur local sub‑tree counts; inter‑graph similarity falls more slowly, so the signal‑to‑noise ratio narrows through $k=7$ and widens again at $k=8$. Effect sizes (Fig~\ref{fig:kernel-stats_cohens}) mirror this dip‑and‑rebound; (Fig~\ref{fig:kernel-stats_p-values}) pinpoints $k=8$ as the first value where the self‑vs‑other distinction regains FDR‑corrected significance while keeping every graph connected.

Guided by these converging cues we fix \textbf{$k=8$} for all subsequent analyses. This density is rich enough to capture long‑range thematic links yet sparse enough to preserve graph distinction, as supported by earlier statistics.

\subsection{Graph‑Level Similarity Landscape}
We surveyed the macroscopic relations among the corpus using this fixed $k=8$.

\begin{figure}
    \centering
    \includegraphics[width=0.67\linewidth]{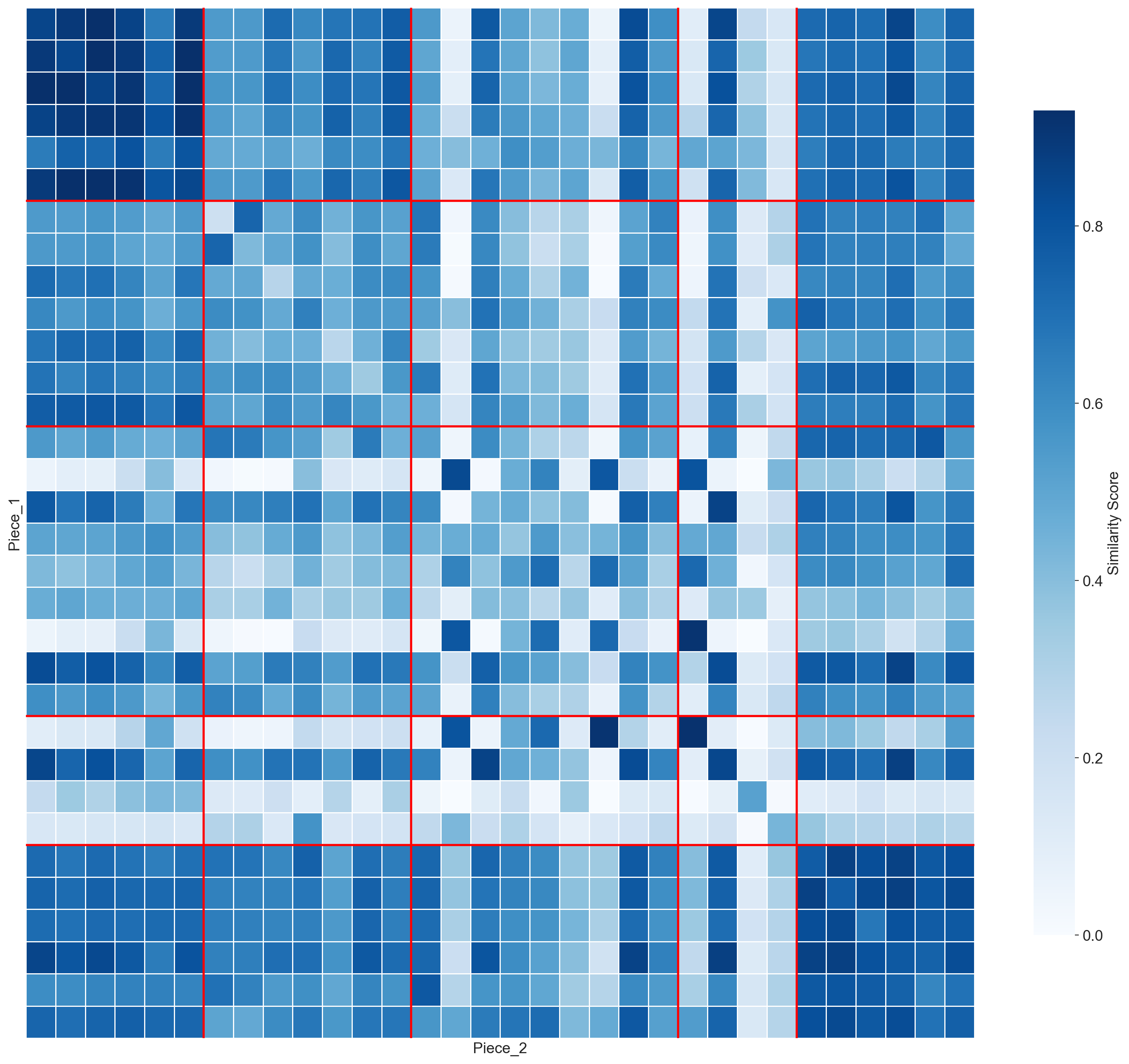}
    \caption{Heatmap of inter-piece similarity. Piece labels are omitted for clarity, and red dividing lines separate compositions by different composers, highlighting grouping patterns indicative of stylistic coherence.}
    \label{fig:grouped_similarity_heatmap}
\end{figure}

\subsubsection{\textbf{Piece‑wise Similarity}}
The heatmap in Fig.~\ref{fig:grouped_similarity_heatmap} reveals prominent intra‑composer cohesion. The darkest diagonal block corresponds to Bach’s Cello Suites, which exhibit extreme internal similarity ($\varphi_{\mathrm{WL}} \geq 0.85$), echoing Winold’s formal and harmonic analyses that describe a tightly unified compositional framework~\cite{winold2007}. Likewise, Chopin’s Études Op.~10 form a strong cluster, particularly Études Nos.~4 and 7, which share virtuosic figuration and motivic density~\cite{moon2011}.

By contrast, Chopin’s Études Op.~25 exhibit markedly lower internal similarity, reflecting a broader expressive and gestural range that prioritizes individualized physical narratives over motivic or harmonic cohesion~\cite{moon2011,montague2012instrumental}. Meanwhile, Ysaÿe’s Violin Sonatas maintain high pairwise similarity, a finding consistent with Rittstieg’s thesis on the Baroque-inspired cohesion of the cycle~\cite{rittstieg2016}.

\begin{figure}
    \centering
    \includegraphics[width=0.8\linewidth]{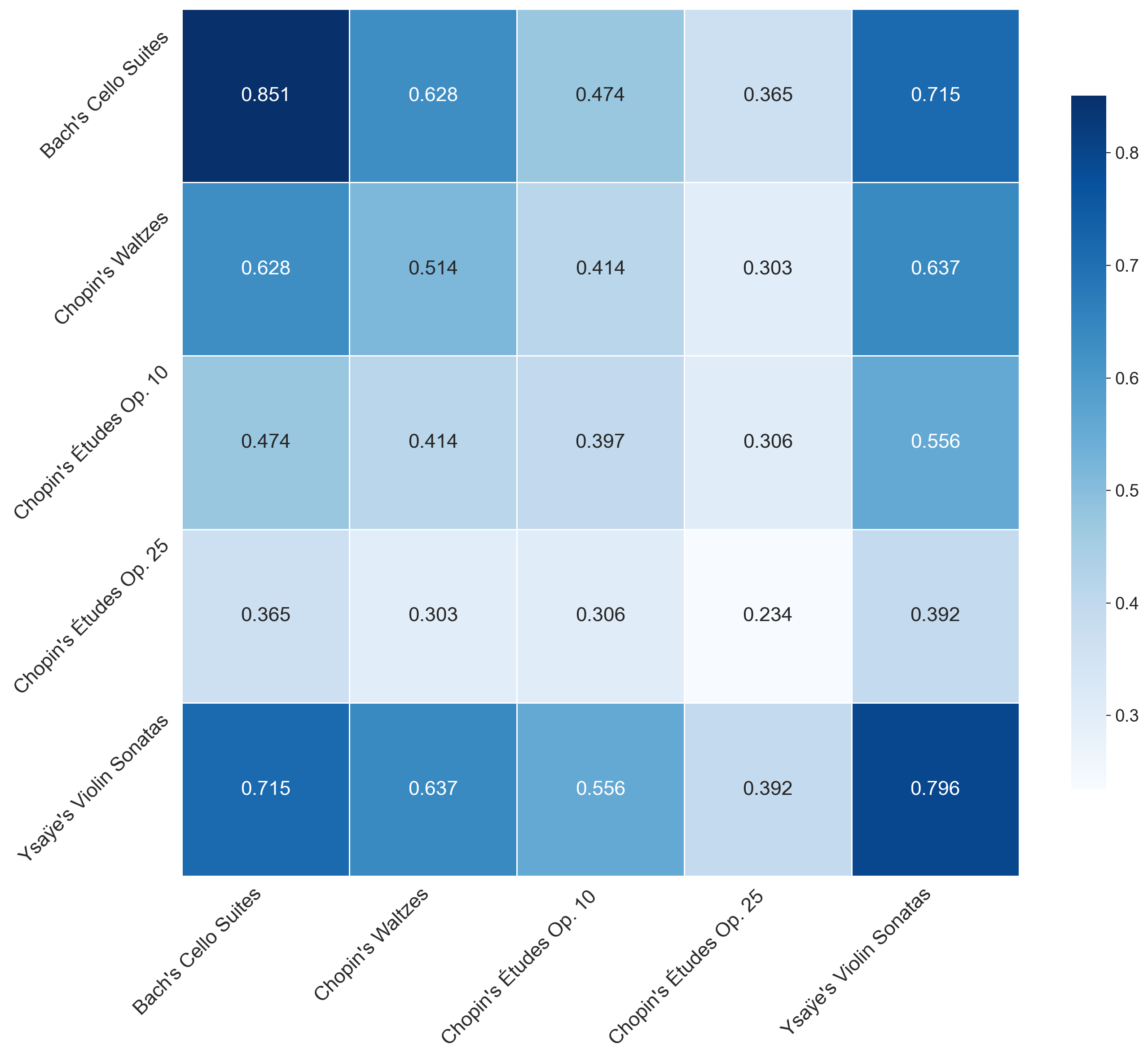}
    \caption{Average similarity matrix for different corpora. Higher values indicate stronger internal structural coherence, while lower values reflect greater divergence between compositions.}
    \label{fig:piece_type_similarity}
\end{figure}

\subsubsection{\textbf{Corpora-Aggregated Similarity}}
Graphs were pooled into five well‑known corpora (Bach Suites, Chopin Waltzes, Chopin Études Op.~10, Chopin Études Op.~25, Ysaÿe Sonatas). Each cell in Fig~\ref{fig:piece_type_similarity} records the mean similarity between every pair of graphs drawn from two corpora.


The diagonal entries confirm strong internal cohesion for Bach’s Cello Suites ($0.85$), Ysaÿe’s Sonatas ($0.80$), and, to a lesser degree, Chopin’s Waltzes ($0.51$). The extremely low value for Chopin’s Op.~25 Études ($0.23$) aligns with prior musicological accounts of their expressive heterogeneity. Cross-corpus similarity remains the highest between Bach and Ysaÿe ($0.71$), supporting Rittstieg’s argument that Ysaÿe’s sonatas were explicitly modeled on Bach’s solo works.

\subsection{Segment‑Level Corroboration via Pairwise MDS}
\begin{figure*}[t]
    \centering
    \begin{subfigure}[t]{0.32\textwidth}
        \includegraphics[width=\linewidth]{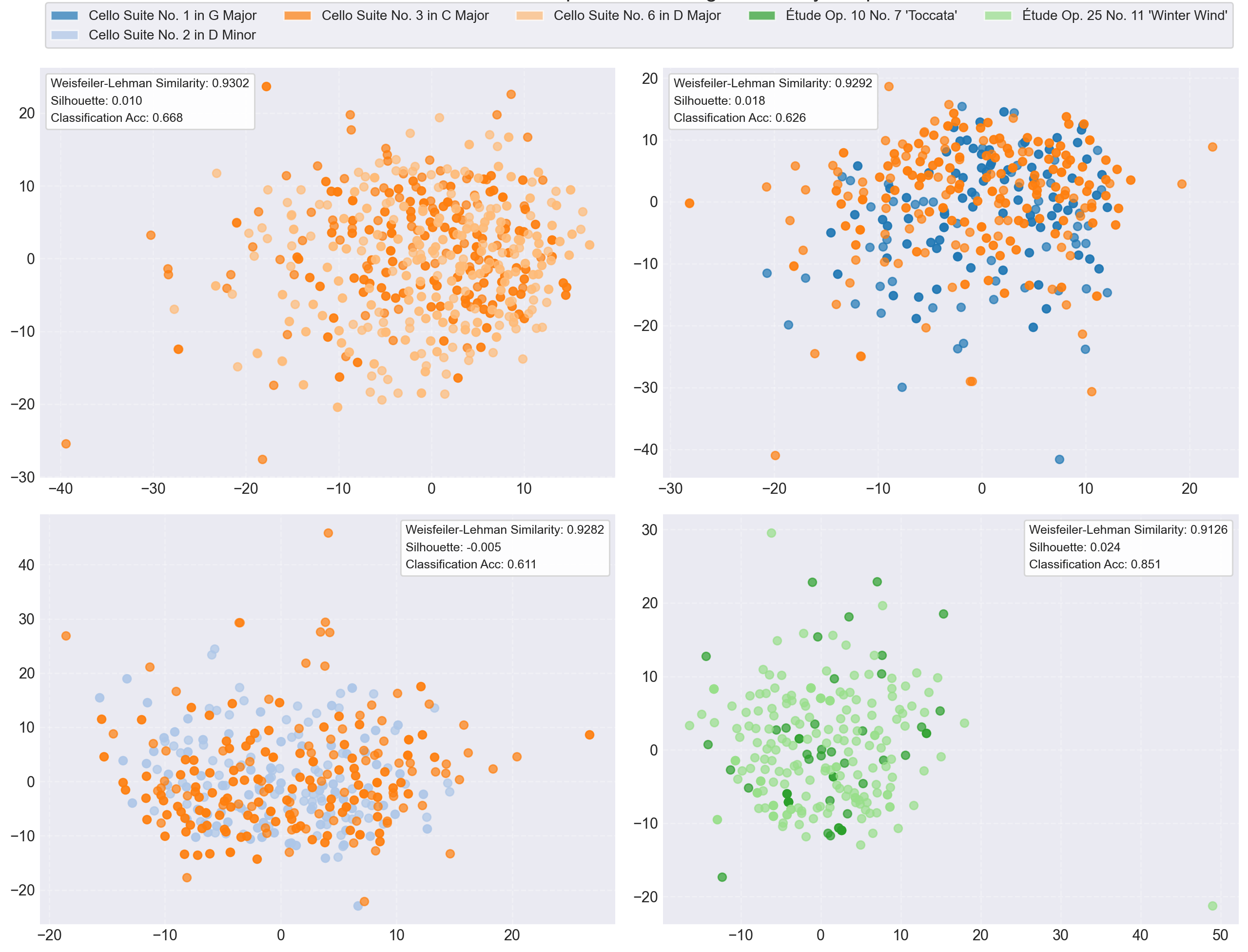}
        \caption{High‑similarity panels}
        \label{fig:mds_high}
    \end{subfigure}
    \hfill
    \begin{subfigure}[t]{0.328\textwidth}
        \includegraphics[width=\linewidth]{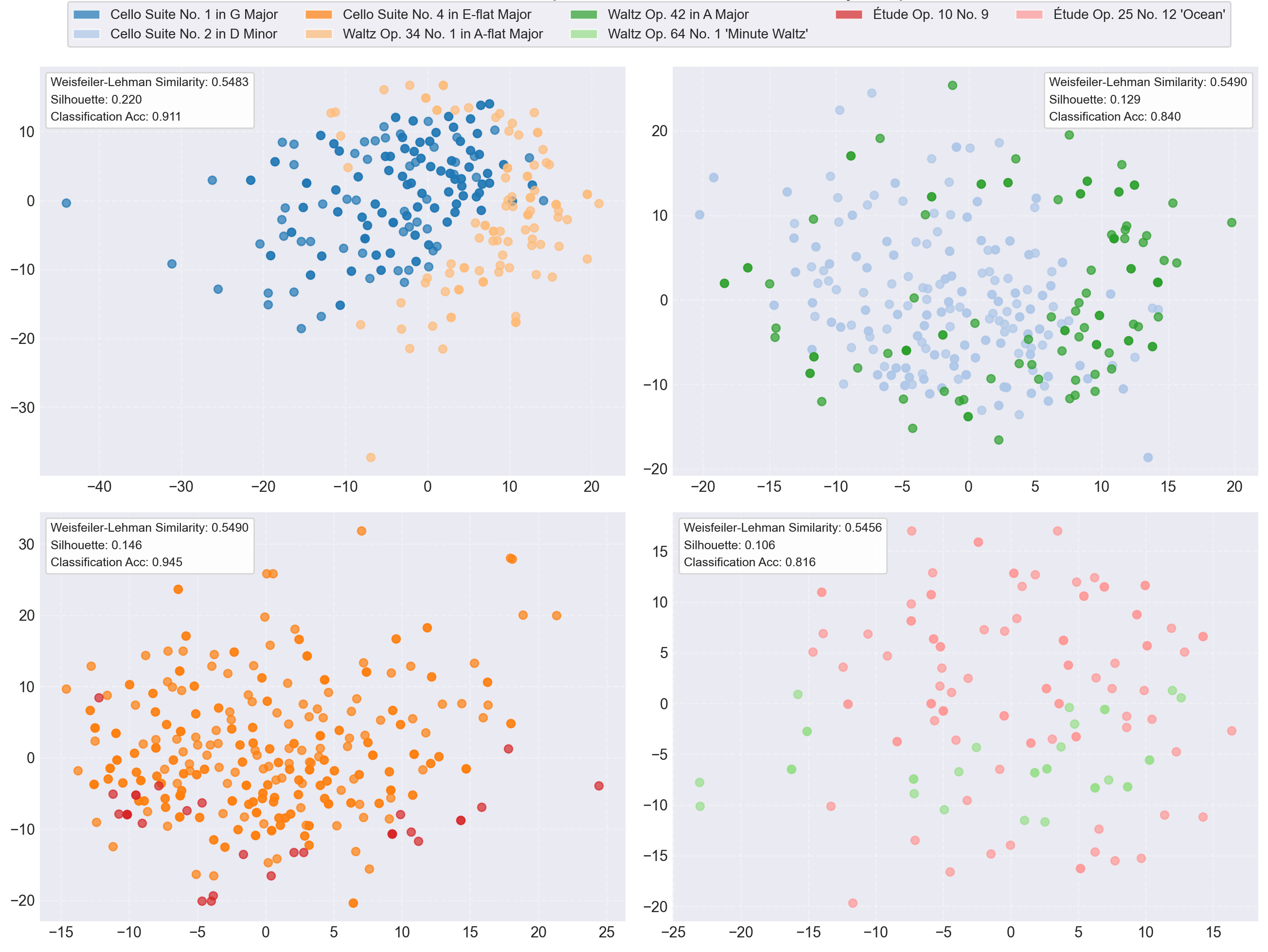}
        \caption{Moderate‑similarity panels}
        \label{fig:mds_medium}
    \end{subfigure}
    \hfill
    \begin{subfigure}[t]{0.32\textwidth}
        \includegraphics[width=\linewidth]{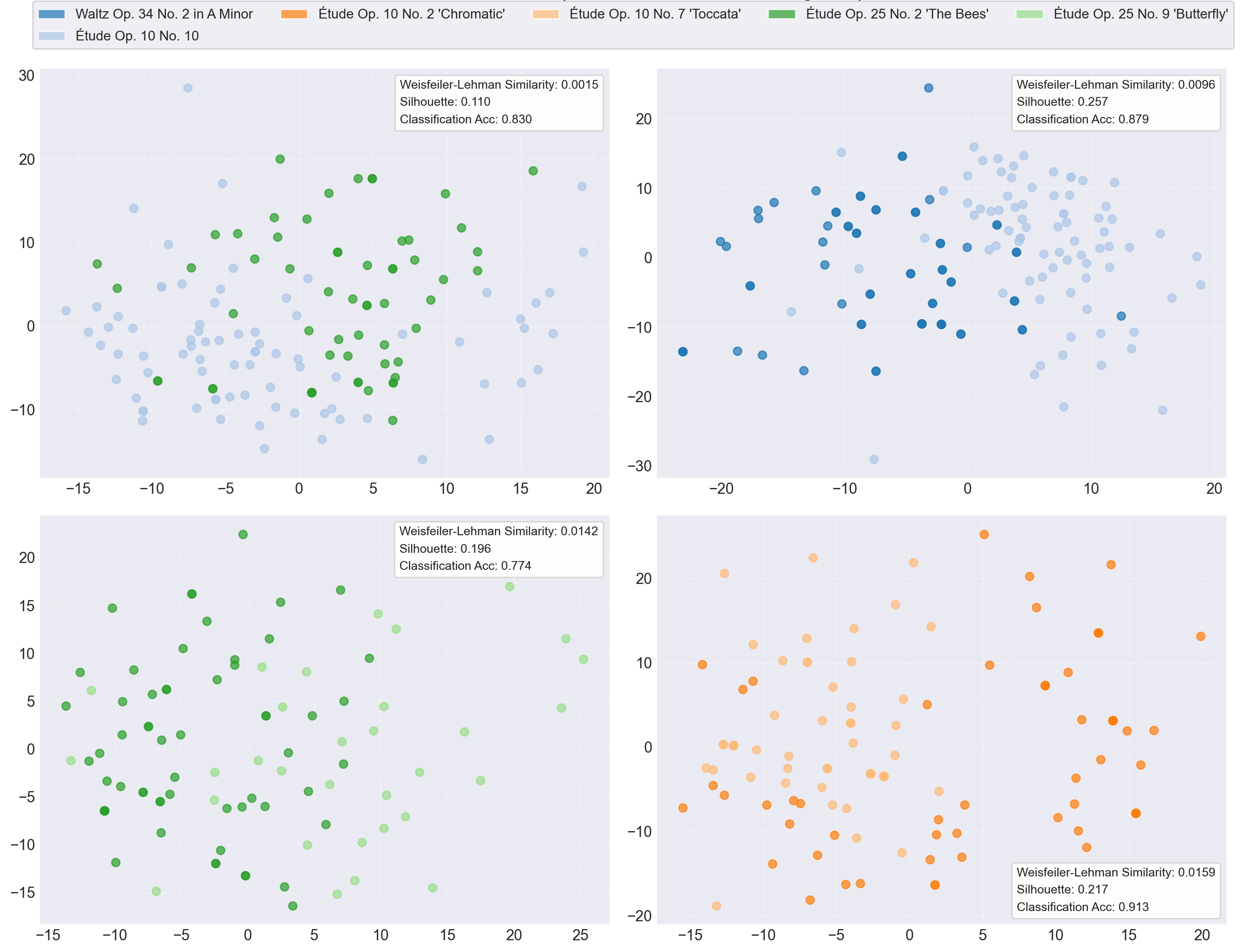}
        \caption{Low‑similarity panels}
        \label{fig:mds_low}
    \end{subfigure}
    \caption{Segment-level MDS embeddings by WL similarity level. Each point is coloured by piece origin, with diagnostic metrics (silhouette $\bar S$, $k$-NN $\bar A$) inset.}
    \label{fig:mds_all}
\end{figure*}

We sampled three sets of four piece pairs—from pieces WL similarities among highest, lowest, and closest to the median—to validate whether pieces that appear similar at the graph-level also occupy overlapping regions at the space of individual segments. Figures can be found at Fig~\ref{fig:mds_all}.

A clear and monotonic correspondence emerges across the three similarity tiers of samples pairwise comparisons:

\begin{itemize}
  \item \textbf{High similarity} is accompanied by very low silhouette (\(\bar S\approx0.01\)) and moderate classification accuracy (\(\bar A\approx0.69\)), indicating that segments from both pieces occupy a largely overlapping manifold.
  \item \textbf{Moderate similarity} yields intermediate separation (\(\bar S\approx0.15\)) and high classification accuracy (\(\bar A\approx0.88\)), reflecting partially overlapping but still distinguishable segment clusters.
  \item \textbf{Low similarity} produces the largest silhouette values (\(\bar S>0.19\)) and robust classification accuracy (\(\bar A>0.85\)), consistent with two clearly disjoint segment manifolds.
\end{itemize}

The segment‐level MDS embeddings provide clear, confirmatory evidence that our WL kernel is capturing genuine local affinities—higher graph‐level similarity corresponds to greater overlap in the raw segment geometry, and low similarity to clear separation.  However, DTW and MDS operates at a more granular level and does not encode how those segments interlock into larger formal arcs; it therefore cannot by itself distinguish pieces whose phrase‐level organisation diverges.

\subsection{Graph Embeddings \& Clustering}
To further validate the distinctiveness of each piece’s graph representation, we generated fixed-length vector embeddings using \texttt{graph2vec} and subsequently performed $k$-means clustering. Cluster assignments are visualized in Figure~\ref{fig:clustering_unlabeled} for each piece, further detailed in Appendix~\ref{appendix:cluster_labels} (Table~\ref{tab:cluster_labels_grouped}).

\begin{figure}[h]
    \centering
    \includegraphics[width=0.6\linewidth]{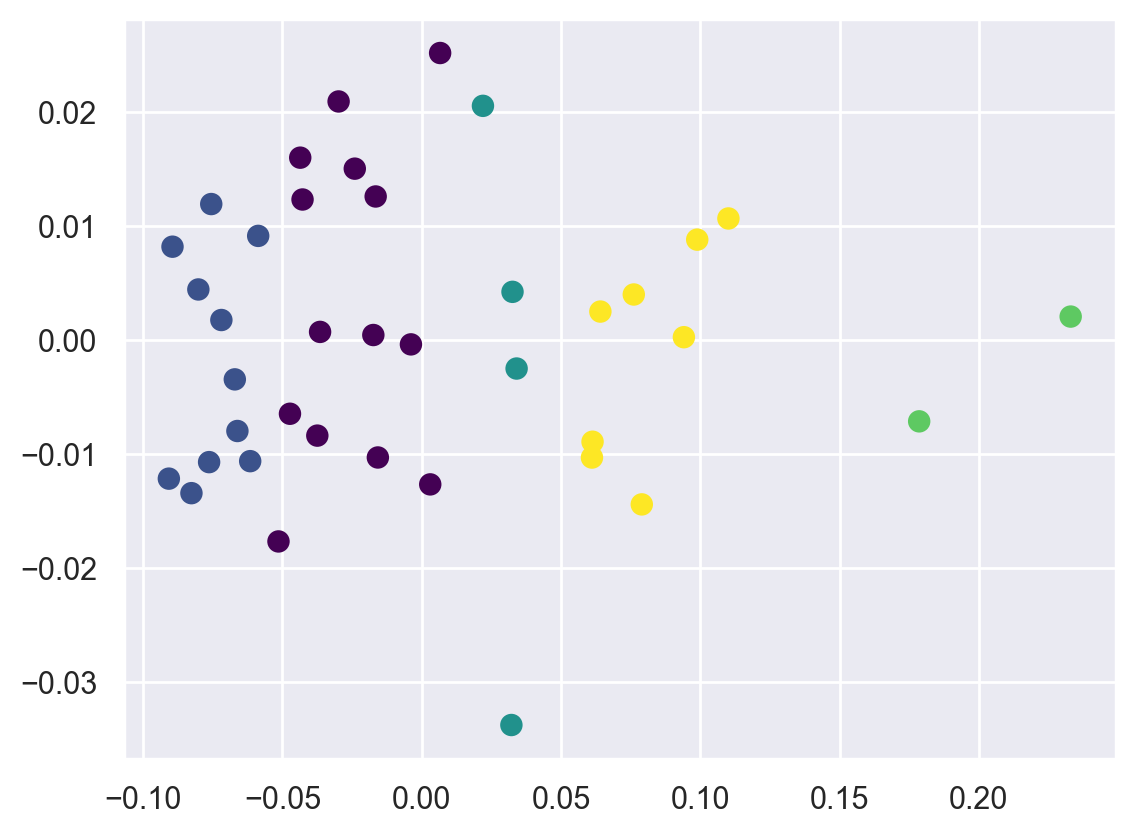}
    \caption{2D scatter plot of the graph embeddings for each piece, color-coded by cluster label. The axes represent the reduced principal components of the embedded space.}
    \label{fig:clustering_unlabeled}
\end{figure}

The clustering results show that compositions tend to group not strictly by composer, but by shared musical traits—such as mood, texture, and structural patterns—indicating that the model captures deeper stylistic and perceptual qualities.

\textbf{Cluster 0} includes several of Chopin’s études and waltzes, which are unified by tightly woven musical ideas and consistent phrasing. These pieces often feature repeated melodic shapes and subtle rhythmic shifts that help organize the music without breaking its flow. Even in freer, more expressive works like Waltz Op. 64 No. 2, Chopin maintains an underlying sense of timing and structure \cite{hood2017interpreting, unlu2024analysis}, which the model likely recognizes as strong internal consistency.

\textbf{Cluster 1} includes Bach’s Cello Suites Nos. 2, 3, and 5, which share a focused, expressive character but differ in their use of melodic space. The Second and Fifth Suites are more introspective, featuring narrow ranges, frequent repetition, and, in the Fifth, distinctive rhythms and tuning \cite{winold2007bach}. In contrast, the Third Suite projects a broader, more open sound through its consistent use of a wide two-octave range. Despite these differences, all three emphasize clear motivic organization and a strong sense of unity, which the model likely captures in their grouping.



\textbf{Cluster 2} contains Bach’s Fourth and Sixth Cello Suites, which are more technically demanding and adventurous. They span a wider pitch range and make strong use of bold, clear patterns based on major chords. The Sixth Suite, in particular, showcases dramatic jumps and fast passages, which suggest a more outward, virtuosic character \cite{winold2007bach}. The model likely clusters these pieces together due to their shared energy, variety, and structural clarity.

\textbf{Cluster 3} links Bach’s Suite No. 1 in G major with several of Ysaÿe’s Violin Sonatas, united by lyrical phrasing, clear formal structure, and Baroque-inspired gestures. While Bach emphasizes graceful melodic turns, Ysaÿe blends similar forms with expressive techniques and nature-inspired imagery \cite{winold2007bach, Yang2016}. Their shared elegance and balance between tradition and innovation likely account for their clustering.

\textbf{Cluster 4} includes Satie’s Gnossiennes, Mozart’s Sonata K. 331, and Chopin’s Étude Op. 10 No. 6 (“Lament”)—all gentler, more introspective works built on repetition, smooth phrasing, and minimal harmonic change \cite{gates2019investigation, simmons2013erik, irving1997mozart}. Despite stylistic differences, their shared calm and structural simplicity likely explain their grouping.


These results suggest that the model captures a range of musical attributes—instrumentation, melodic contour, rhythmic articulation, and expressive pacing—beyond composer identity. Graph-based representations thus reveal not only intuitive groupings but also subtler, cognitively grounded relationships among stylistically diverse compositions.

\section{Conclusions}

This study presents a graph-based method for representing classical compositions by combining cognitive music theory (IR and Temporal Gestalt) with computational tools like DTW, graph kernels, MDS, and embeddings. By structuring melodies into perceptual segments within k-NN graphs, the approach effectively captures distinctive musical identities across compositions.

Intra- and inter-graph similarity comparisons using the Weisfeiler-Lehman kernel revealed that segment graphs exhibit significantly higher internal coherence than cross-compositional similarity, validating their ability to encode musical identity. Segment-level analyses using DTW and MDS further corroborated these findings, showing that graph similarity aligns with local structural overlap in segment geometry. Graph2vec embeddings and clustering confirmed that these representations capture stylistic and structural traits that go beyond mere composer identity, often grouping pieces by shared expressive or formal features.

These results highlight the potential of graph-based methods for structured, perceptually grounded music analysis. Future work could scale this approach to larger, more diverse datasets, explore alternative similarity measures, and incorporate broader musical features or traditions, while also improving computational efficiency.

\appendix
\label{appendix:cluster_labels}
\begin{table*}[ht]
    \centering
    \footnotesize
    \caption{Cluster assignments from the $k$-means algorithm.}
    \label{tab:cluster_labels_grouped}
    \begin{tabular}{l p{0.35\linewidth} c | l p{0.35\linewidth} c}
    \toprule
    \textbf{Composer} & \textbf{Piece Title} & \textbf{Label} 
      & \textbf{Composer} & \textbf{Piece Title} & \textbf{Label} \\
    \midrule
    \multicolumn{6}{l}{\textbf{Cluster 0}} \\
    \midrule
    Chopin & Waltz Op.\ 34 No.\ 2 in A Minor                           & 0 
          & Chopin & Waltz Op.\ 64 No.\ 2 in C Minor                       & 0 \\
    Chopin & Waltz Op.\ 64 No.\ 3 in A\textminus flat Major             & 0 
          & Chopin & Étude Op.\ 10 No.\ 1 `Waterfall'                       & 0 \\
    Chopin & Étude Op.\ 10 No.\ 10                                      & 0 
          & Chopin & Étude Op.\ 10 No.\ 2 `Chromatic'                       & 0 \\
    Chopin & Étude Op.\ 10 No.\ 4 `Torrent'                              & 0 
          & Chopin & Étude Op.\ 10 No.\ 5 `Black Keys'                      & 0 \\
    Chopin & Étude Op.\ 10 No.\ 7 `Toccata'                              & 0 
          & Chopin & Étude Op.\ 10 No.\ 8 `Sunshine'                        & 0 \\
    Chopin & Étude Op.\ 25 No.\ 12 `Ocean'                              & 0 
          & Chopin & Étude Op.\ 25 No.\ 9 `Butterfly'                       & 0 \\
    Mozart & Piano Sonata No.\ 10 in C Major, K.\ 330 (1st movement)    & 0 
          & Ysa\"{y}e & Violin Sonata No.\ 3 in D Minor                      & 0 \\
    \midrule
    \multicolumn{6}{l}{\textbf{Cluster 1}} \\
    \midrule
    Bach   & Cello Suite No.\ 2 in D Minor                              & 1 
          & Bach   & Cello Suite No.\ 3 in C Major                          & 1 \\
    Bach   & Cello Suite No.\ 5 in C Minor                              & 1 
          & Chopin & Waltz Op.\ 42 in A Major                               & 1 \\
    Ysa\"{y}e & Violin Sonata No.\ 1 in G Minor                          & 1 
          &        &                                                       &   \\
    \midrule
    \multicolumn{6}{l}{\textbf{Cluster 2}} \\
    \midrule
    Bach   & Cello Suite No.\ 4 in E\textminus flat Major              & 2 
          & Bach   & Cello Suite No.\ 6 in D Major                          & 2 \\
    \midrule
    \multicolumn{6}{l}{\textbf{Cluster 3}} \\
    \midrule
    Bach   & Cello Suite No.\ 1 in G Major                              & 3 
          & Chopin & Waltz Op.\ 34 No.\ 1 in A\textminus flat Major         & 3 \\
    Chopin & Étude Op.\ 25 No.\ 11 `Winter Wind'                        & 3 
          & Mozart & Minuet in G Major                                      & 3 \\
    Ysa\"{y}e & Violin Sonata No.\ 2 in A Minor                          & 3 
          & Ysa\"{y}e & Violin Sonata No.\ 4 in E Minor                       & 3 \\
    Ysa\"{y}e & Violin Sonata No.\ 5 in G Major                          & 3 
          & Ysa\"{y}e & Violin Sonata No.\ 6 in E Major                       & 3 \\
    \midrule
    \multicolumn{6}{l}{\textbf{Cluster 4}} \\
    \midrule
    Chopin & Waltz Op.\ 34 No.\ 3 in F Major                            & 4 
          & Chopin & Waltz Op.\ 64 No.\ 1 `Minute Waltz'                    & 4 \\
    Chopin & Étude Op.\ 10 No.\ 6 `Lament'                              & 4 
          & Chopin & Étude Op.\ 10 No.\ 9                                     & 4 \\
    Chopin & Étude Op.\ 25 No.\ 2 `The Bees'                            & 4 
          & Mozart & Piano Sonata No.\ 11 in A Major, K.\ 331 `Rondo Alla Turca' & 4 \\
    Satie  & Gnossienne No.\ 1                                          & 4 
          & Satie  & Gnossienne No.\ 2                                       & 4 \\
    Satie  & Gnossienne No.\ 3                                          & 4 
          & Satie  & Gnossienne No.\ 4                                       & 4 \\
    \bottomrule
    \end{tabular}
\end{table*}

\section{Computational Efficiency}  
To manage the high time complexity of DTW (\(O((mn)^2)\) for full pairwise comparisons), we use multiprocessing and a checkpoint system. Segment pairs are processed in parallel and saved in chunks using Python’s pickle module, allowing interrupted computations to resume without redundancy.

\bibliographystyle{ACM-Reference-Format}
\bibliography{references}

\end{document}